\documentclass[a4paper,twoside]{article}
\newcommand{\affil}[1]{$^{\rm #1}$}

\usepackage[authoryear]{natbib}
\bibpunct{(}{)}{;}{a}{}{,}
\usepackage{amsmath}
\usepackage{graphicx}
\usepackage[english]{babel}

\date{}

\title{\large\bf\flushleft The Radio-FIR correlation in the Milky Way}
\author {
   {J. Zhang\affil{A}, A.Hopkins\affil{A,B}, P.J. Barnes\affil{A,C}, M. Cagnes\affil{A}, Y. Yonekura\affil{D}, Y. Fukui\affil{E}} \\
{\small \affil{A}\,Sydney Institute for Astronomy, School of Physics, University of Sydney, NSW 2006, Australia}\\
{\small \affil{B}\,Anglo-Australian Observatory, PO Box 296, Epping, NSW 1710, Australia}\\
{\small \affil{C}\,Astronomy Department, University of Florida, Gainesville,
FL 32611, USA}\\
{\small \affil{D}\,Faculty of Science, Ibaraki University, 2-1-1 Bunkyo, Mito,
Ibaraki 310-8512, Japan}\\
{\small \affil{E}\,Department of Astrophysics, Nagoya University, Furo-cho,
Chikusa-ku, Nagoya 464-8602, Japan}
}

\begin{document}

\twocolumn

\maketitle

\small{\bf Abstract: We investigate the scale on which the correlation  arises between the $843\,$MHz
radio and the $60\,\mu$m far-infrared (FIR) emission from star forming regions in the Milky way. The correlation, which exists on the smallest scales investigated (down to $\approx\,4\,$pc), becomes
noticeably tight on fields of size $30'$, corresponding to physical scales of $\approx\,20-50\,$pc.
The FIR to radio flux ratio on this scale is consistent with the radio emission being dominated by thermal emission. We
also investigate the location dependence of $q_{\rm mean}$, a parameter measuring the mean FIR
to radio flux ratio, of a sample of star forming regions. We show that $q_{\rm mean}$ displays a
modest dependence on galactic latitude. If this is interpreted as a dependence on the intensity of
star formation activity, the result is consistent with studies of the Large Magellanic Cloud (LMC)
and other nearby galaxies that show elevated values for $q$ in regions of enhanced star formation.}

\section{Introduction}
The tight correlation between radio and far-infrared (FIR) emission in external galaxies has been long
known \cite[e.g.,][]{Yun:01}. To explain the correlation, the conventional picture suggests
that both emissions are the result of massive star formation. The radio emission has both a thermal
component and a non-thermal synchrotron component due to cosmic ray electrons accelerated to
relativistic speeds through supernovae shocks. The FIR is mostly thermal emission from
dust excited by UV emission associated with star formation. The FIR emission is a common measure of
star formation rates \cite[SFRs,][]{Ken:98}, although a contribution to the FIR emission from
older stellar populations \citep{Bel:03} necessitates a more complex calibration from luminosity
to SFR than a simple linear relation. The effects of different dust geometries and opacities in the
immediate vicinity of different star forming regions potentially adds a significant complication
\citep{Cal:97,Cal:95}. There is also a strong trend for higher levels
of dust obscuration to occur preferentially in galaxies of higher SFR \cite[e.g.,][]{Per:03,Afo:03,Hop:01},
so it is not unreasonable to expect a similar variation in individual star forming regions.

A complication in understanding the radio-FIR correlation arises from the fact that synchrotron
radiation occurs on a larger spatial scale, due to its diffusion length, than the FIR thermal emission,
\cite[see, e.g.,][]{Mur:06}. Measurements of the total FIR to radio flux ratio, $q$ \citep{Hel:85},
for the Milky Way \citep{Bro:89} show that on the largest scales our Galaxy is consistent with other
star forming galaxies, although these authors also emphasise that the tight global correlation is not
seen within galaxies. Measurements of $q$ within a few hundred pc around Orion
\citep{Bou:88} give a value elevated compared to the average for star forming galaxies, and similar
to those seen in regions of active star formation in the Large Magellanic Cloud (LMC) by \cite{Hug:06}
and within nearby galaxies \citep{Mur:06}. A question of interest then becomes, on what
spatial scale does the radio-FIR correlation begin to appear around star forming regions? It has been proposed that
this should start to appear on scales comparable to the $\approx 1-2\,$kpc diffusion length for cosmic
ray electrons \citep{Bic:90,MH:98}. This would correspond to the observed region around an area of
massive star formation being large enough to include sufficient quantities of
radio synchrotron emission. We show that, for the Milky Way, a radio-FIR correlation exists on
much smaller scales, and is likely to be a correlation between the thermal radio and FIR components.

Recent studies of the radio-FIR correlation have concentrated on analysing the physical processes which lead to the correlation on galactic scales. This has been done by analysing the correlation on small spatial scales \citep{Xu:92,Hug:06}, and within other nearby galaxies \citep{Mur:06}.
\cite{Hug:06}, looking at the Large Magellanic Cloud (LMC), achieved a
spatial resolution of $20\,$pc.
On such fine spatial scales, these authors have demonstrated varying $q$ for different structures in
the LMC. In regions of the most active star formation, $q$ is higher. This is interpreted as due to two
different physical mechanisms driving the radio-FIR correlation. In the most active star forming regions,
radio emission is mostly thermal while in less active regions, the correlation is due to synchrotron
radiation. We compute the mean $q$, called $q_{\rm mean}$ herein, for star forming regions
grouped by their galactic latitude. We show that $q_{\rm mean}$ is elevated closer to the galactic
plane, where the most active star formation in our galaxy is occuring, compared to the value at higher galactic latitudes.

\section{Analysis}
We investigate the spatial scale on which the radio-FIR correlation begins to arise around massive
star forming regions in the Milky Way. We then extend this analysis to explore whether the radio-FIR
correlation shows any Galactic latitude dependence, as might be expected if $q$ depends on
star formation rate, as suggested by the elevated $q$ values seen in the LMC bar \citep{Hug:06}
and spiral arms of nearby galaxies \citep{Mur:06}.

\subsection{Data}
The star forming regions used for this analysis are drawn from the Census of
High and Medium-mass Protostars (CHaMP) survey conducted with the Mopra dish of the Australia Telescope\footnote{The Mopra telescope is part of the Australia Telescope which is funded by the
Commonwealth of Australia for operation as a National Facility managed by CSIRO. The University of
New South Wales Digital Filter Bank used for the observations with the Mopra Telescope was provided
with support from the Australian Research Council.}
\citep{Bar:06,Bar:09a}. CHaMP targeted massive dense gas clumps based on the Nanten Survey
\citep{Yon:05} of carbon monoxide gas, a standard tracer of the sites of star formation. The CHaMP
catalogue is based on an examination of complete Nanten maps for the region
$280^{\circ} < l < 300^{\circ}$, $-4^{\circ} < b < +2^{\circ}$, made in four $J=1-0$ molecular spectral
lines in the 3\,mm band: CO, $^{13}$CO, C$^{18}$O, and HCO$^+$ \cite[see also][for a
subset of these data]{Yon:05}. The CHaMP catalogue contains 209 local emission maxima in
$(l,b,v)$ space, identified from the Nanten C$^{18}$O and HCO$^+$ data cubes \citep{Bar:09b}. In this
sense, the CHaMP catalogue comprises an unbiased sample of all the dense gas clumps in this
window, and is therefore a useful indicator of the most likely sites of current massive star formation in
this part of the Milky Way. The unbiased nature of the CHaMP selection, allowing identification of all
massive star forming regions throughout a representative spiral arm of the Milky Way, suggests that
our analysis should be generally applicable to the Galaxy as a whole.

		\begin{figure*} [th]
		\begin{center}
		\includegraphics[scale=0.50,keepaspectratio]{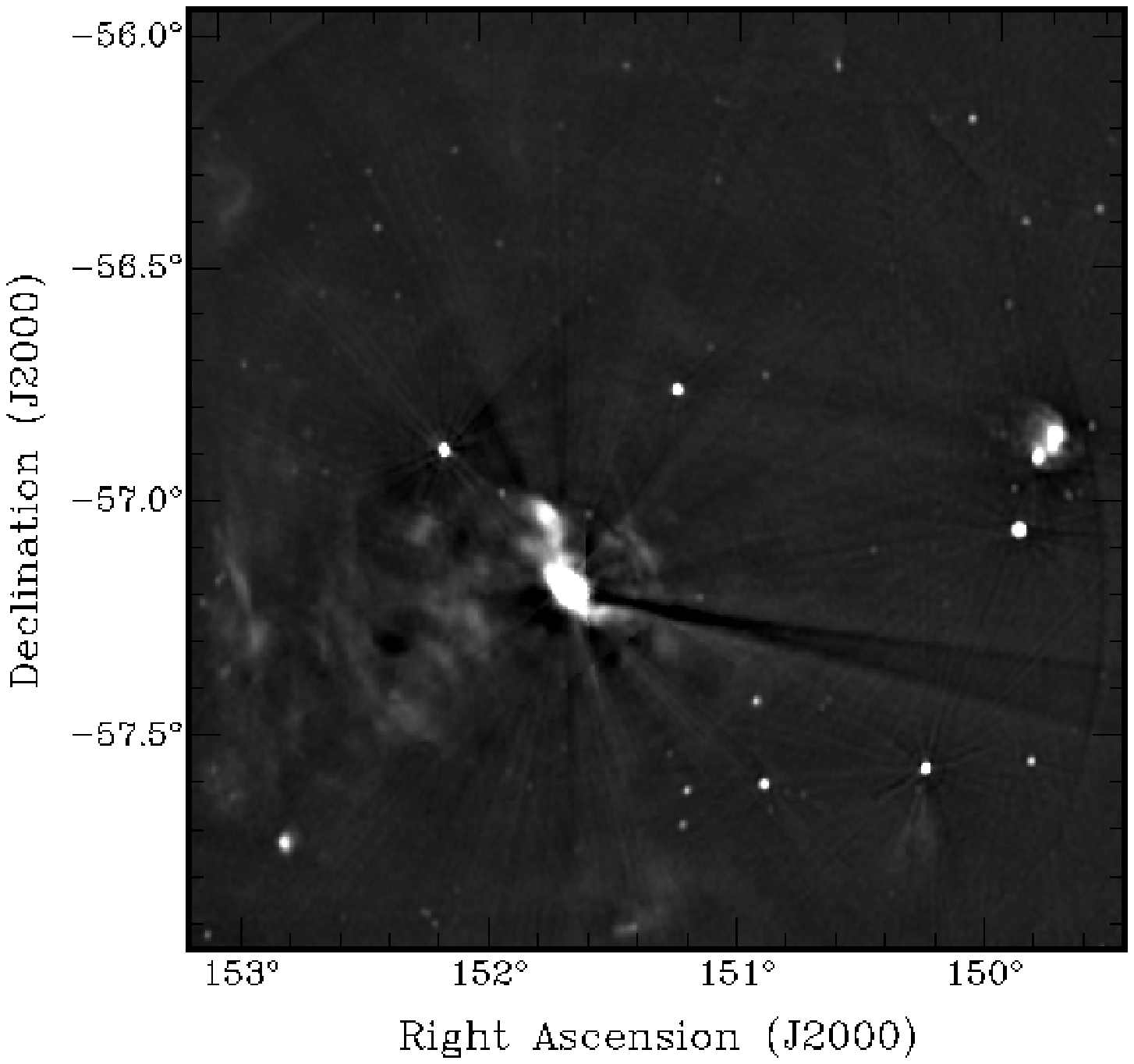}
		\hfill
	         \includegraphics[scale=0.40,keepaspectratio]{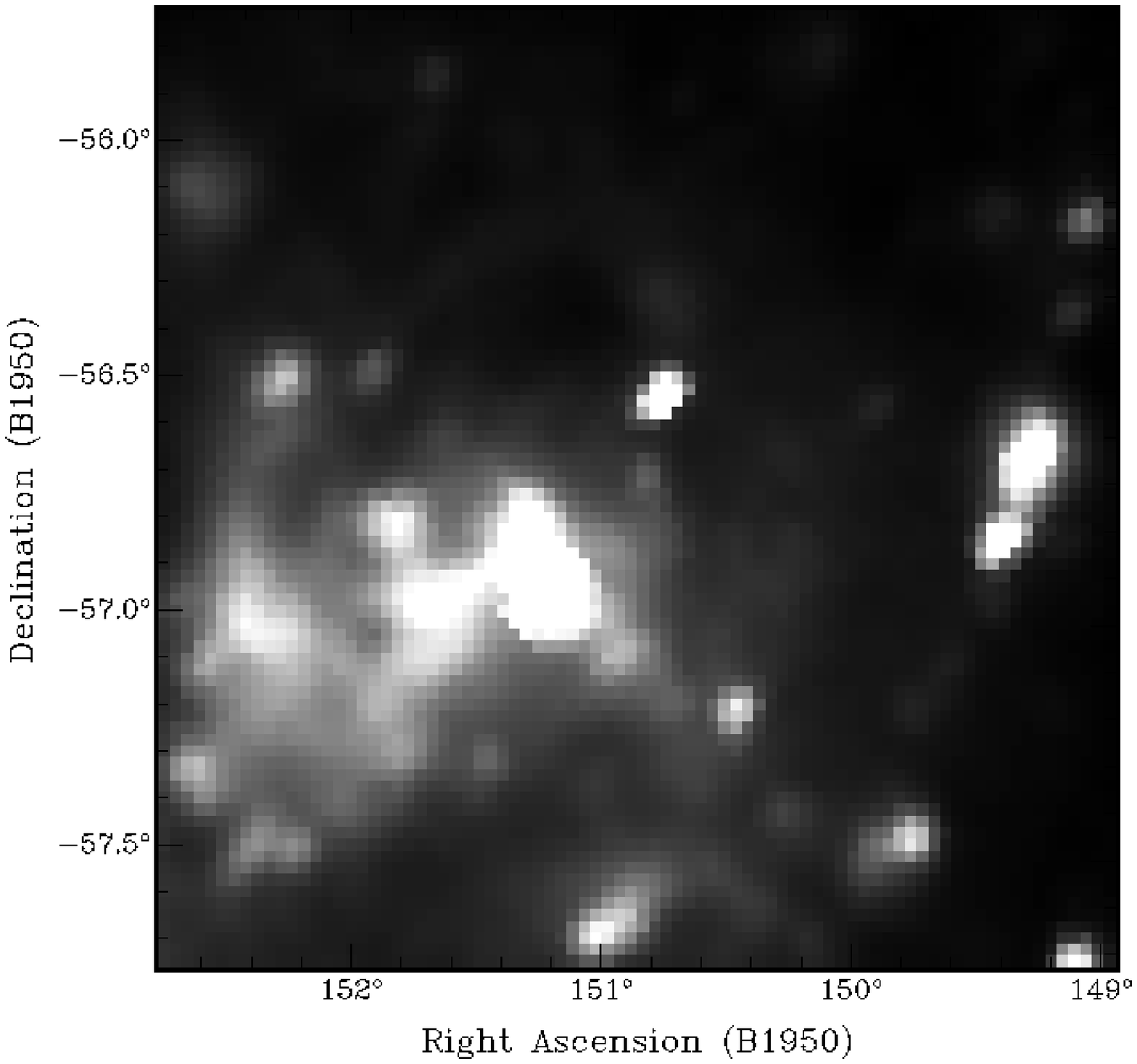}
		\caption {SUMSS 843\,MHz (left) and IRAS $60\,\mu$m (right) images for a
		$2^{\circ}\times2^{\circ}$ region centred on one of the target CHaMP massive star
		forming regions.
		\label{fig1}}
		\end{center}
		\end {figure*}

Using the coordinates of the 173 highest CO luminosity sources from the CHaMP survey, being
those already catalogued at the initiation of this study, 138 target
regions were selected for this analysis based on availability of both radio and FIR images. The
omission of the $20\%$ of targets lacking radio
and FIR data should not introduce any significant bias to our analysis, as we still sample the full
range of Galactic latitude and longitude within the CHaMP survey area. The radio images at
$843\,$MHz were obtained from the Sydney University Molonglo Sky Survey
\cite[SUMSS,][]{Boc:99,Mau:03} and FIR images at $60\mu$m from the Infrared Astronomical
Satellite (IRAS) Sky Survey Atlas (ISSA)\footnote{This research has
made use of the NASA/ IPAC Infrared Science Archive, which is operated by the Jet Propulsion
Laboratory, California Institute of Technology, under contract with the National Aeronautics and Space
Administration.}. An advantage of the SUMSS radio images for this analysis is the high
sensitivity to extended emission, a consequence of the closely-spaced and highly redundant
spacings between receivers in the Molonglo Observatory Synthesis Telescope \citep{Boc:99}.
The resolution of the SUMSS images is $45" \times 45"$cosec$|\delta|$, and the internal flux
density scale is accurate to within $3\%$ \citep{Mau:03}. An example of both SUMSS and IRAS
images is given in Figure~\ref{fig1}, showing a region two degrees on a side, centred on one
of the CHaMP targets.

\subsection{The spatial scale on which the correlation arises}
\begin{figure*} [th]
		\begin{center}
		\includegraphics[angle=-90,scale=0.60,keepaspectratio]{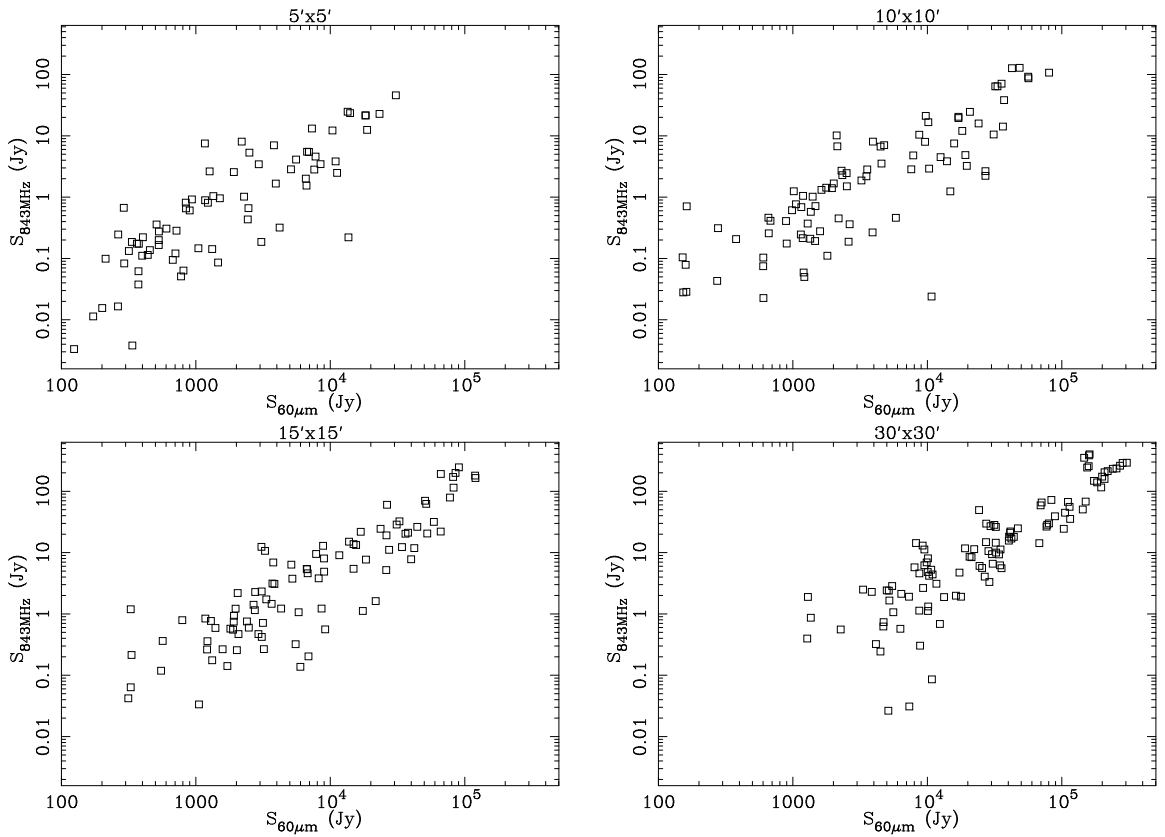}
		\caption {SUMSS 843\,MHz radio flux density as a function of IRAS $60\,\mu$m flux, for
		regions of progressively larger size centred on CHaMP massive star forming regions.
		The radio-FIR correlation tightens noticeably once the area analysed reaches about
		$30'\times30'$. This corresponds to a physical scale of $\approx$ 20\,pc to 50\,pc.
		\label{fig2}}
		\end{center}
		\end {figure*}

For each CHaMP star forming region, SUMSS and ISSA images of varying sizes were obtained.
Square field sizes of $5'$, $10'$, $15'$, $30'$, and $1^{\circ}$ were analysed. Regions $2^{\circ}$
on a side were also obtained, but most of the SUMSS images of this size were incomplete. This
is a limitation of the SUMSS image server, which does not mosaic adjacent images
if the requested field extends beyond the boundary of the nearest $4^{\circ} \times 4^{\circ}$
archival image.
To measure the correlation between the FIR and radio images of the same star forming region, a simple
flux sum was taken by adding up the amplitudes of all pixels within the image. Figure~\ref{fig2}
shows radio flux density as a function of FIR flux for the four smallest spatial scales measured. Each of
the 138 star forming regions is shown, except for those fields where the flux summed to a negative
value in the radio images due to the presence of artifacts. Such fields were not included in any of
our analyses.

The correlation coefficient was calculated for each field size (Table~\ref{tab1}). A significant correlation
is found for all, even the smallest field sizes. By fields of size $30'\times30'$ this correlation becomes
visibly tight, especially for the brightest sources (associated with high SFRs). The correlation at the smallest scales ($5'$ to $30'$) was tested using FIR images randomly offset from the radio images.
The correlation coefficients found for these non-associated regions are significantly reduced, as would be expected if the correlation
is due to the emission from the massive star forming regions. For the randomly offset fields, the
correlation coefficient varies from 0.61 for the $5'$ fields
to 0.64 at $15'$, increasing slightly to 0.75 at $30'$. This supports the suggestion that, even on
the smallest scales measured here, there is a measurable radio-FIR correlation associated with
the CHaMP massive star forming regions.
A value for $q_{\rm mean}$ was calculated, along with the corresponding rms scatter in the
individual $q$ values for each field.
The observed rms scatter in $q$ for our targets is consistent with the ranges of $q$ found within
galaxies by \cite{Hug:06} and \cite{Mur:06}. It is likely that much of this observed scatter is real and
not simply a result of limitations
in our approach, or due to measurement uncertainty. The carbon monoxide isotopologues, used
as a tracer by the CHaMP survey, identify the locations of the coldest and densest gas clumps.
Such dense gas clumps are variously associated with pre-stellar gas, active star forming regions,
and residual clouds from recent massive star formation. Such variation could account for a
significant portion of the scatter observed in $q$.

		\begin{table}
		\begin{center}
		  \begin{tabular}{ c  c  c  c }
		    \hline
		    field size & correlation & $q_{\rm mean}$ & rms  \\
			(arcmin) & coefficient & &	scatter 	\\ \hline 
			5 & 0.86 & 3.41 & 0.54	\\ 
			10 & 0.83 & 3.34 & 0.54 	\\ 
			15 & 0.87 & 3.33 & 0.47	\\ 
			30 & 0.87 & 3.42 & 0.48	\\ 
			60 & 0.81 & 3.56 & 0.44	\\ 
		    \hline 
		  \end{tabular}
		\caption{The correlation between the FIR and radio emission around regions identified in the
		CHaMP survey in the Milky Way. Values of
		$q_{\rm mean}$ and the rms scatter in individual $q$ values are also shown.
                  \label{tab1}}
		\end{center}
		\end{table}

In order to ensure these measurements provide robust estimates of $q_{\rm mean}$, we explored
several possible sources of bias: (1) The presence of artifacts in the radio images; (2) The contribution
of a non-zero background level in the ISSA images; (3) Overlapping fields, arising from the clustering
of CHaMP sources. For each of these tests the values of $q$ and $q_{\rm mean}$
were remeasured using the $30'\times 30'$ images, and the correlation coefficient
recalculated to ensure no bias was introduced. The initial correlation coefficient found for
the $30'\times 30'$ images was 0.87. 
Images removed in the artifact analysis include those displaying radial artifacts originating from
bright sources, and those with large negative-flux regions associated with bright sources, along with
images that did not encompass the full field of view as a consequence of the SUMSS image
archive limitations mentioned above. Excluding those radio images visually
identified as containing artifacts marginally decreases the correlation coefficient to 0.84. It was noted
that the rms scatter of $q$ in the sample with visually identified artifacts removed was not
significantly changed, indicating that such artifacts are unlikely to be the primary source of the
observed scatter.
The median background level in each IRAS image was measured from the histogram of pixel
intensities, and subtracted prior to calculating $q_{\rm mean}$. This resulted in only minor
changes to the resulting $q$ values as the background level was low compared to the total flux.
The resulting correlation coefficient is 0.85. To identify the impact of image overlap,
$q_{\rm mean}$ was calculated only
for images independent by more than 75\% of their size. The resulting correlation coefficient
is 0.74, possibly suggesting that overlapping fields account for some of the strong correlation seen
in Table~\ref{tab1}. A smaller fraction of the $15'$ fields, though, are similarly overlapping,
and removing those does not significantly change the correlation coefficient found
(0.86) compared to the full sample (0.87) for that field size. We conclude that our estimate of the
field size on which the radio-FIR correlation arises is not sensitive to these limitations.

To infer the spatial scale associated with this angular size we require the distances to the
CHaMP star forming regions. Although there are currently only a limited number of distances available,
we make use of those in hand to estimate the length scale on which the radio-FIR correlation
arises. For
our sample of 138 star forming regions, there are 54 distance measurements, based on a combination
of techniques. Some of the objects have distance estimates based on nearby or embedded
HII regions, while
most are estimated kinematically. The $(l,v)$ diagrams in the direction of our sample fields are
largely unambiguous, with distance uncertainties of $\approx 10-20\%$. Less than 50\% of
the distance measurements suffer from a near/far ambiguity. Using these estimates, we can
identify the approximate range on which the radio-FIR correlation begins. An angular
scale of $30'$ corresponds to a spatial scale of 22\,pc to 59\,pc at the estimated distances
of 2.5\,kpc to 6.8\,kpc. This reflects an approximate diameter centred on the star forming regions.
The $5'$ field sizes, on which a correlation also exists, correspond to (diameter) scales
of 3.6\,pc to 9.9\,pc. These scales are all significantly smaller than the diffusion length scale expected
for cosmic ray electrons \citep{Bic:90}, and support the idea that correlations between
different components of the radio and FIR emission exist over a variety of spatial scales. This result
will be explored further in the discussion. In the next section we quantify our measurement
of $q_{\rm mean}$ and explore its galactic latitude dependence.

\subsection {Galactic latitude dependence}
The average $60\mu$m FIR to $843$ MHz radio ratio of selected images is measured by the
$q_{\rm mean}$ parameter and is defined as:
	\begin{equation}
	  q_{\rm mean} = \langle \log_{10}\left(\frac{S_{\rm 60\mu m}}{S_{\rm 843MHz}}\right) \rangle .
	\end{equation}
Note that this differs from the $q$ typically defined in that here we use different FIR wavelength and
radio frequency. This definition of $q_{\rm mean}$ will not affect the trends observed. The use of
843\,MHz rather than 1.4\,GHz radio measurements will act to reduce our $q_{\rm mean}$ by up to
0.15 (for pure synchrotron emission) compared to that of \cite{Hug:06}. This estimate simply
assumes a conversion from 843\,MHz to 1.4\,GHz using a spectral index $\alpha=-0.7$
(with $S\propto\nu^{\alpha}$). Using the $60\,\mu$m flux
only, rather than the combined $60\,\mu$m and $100\,\mu$m FIR flux reduces our estimate of
$q_{\rm mean}$ by approximately 0.15 again, estimated assuming that on average
$S_{100\mu}\approx 1.6\,S_{\rm 60\mu m}$, and using the definition of FIR of \cite{Hel:85}.
This total systematic offset of $\approx 0.3$ is slightly smaller than the rms
scatter in our measurement (Table~\ref{tab1}).

Values for $q_{\rm mean}$ are shown in Table~\ref{tab1}. After correcting the FIR
images by subtracting the background level, we find $q_{\rm mean}=3.34$ for the $30'$ fields,
and $q_{\rm mean}=3.50$ for the $1^{\circ}$ fields, only marginally smaller than
the initial estimates, given the observed scatter. The values that we find here for $q_{\rm mean}$,
especially after considering the systematic offset of $\approx 0.3$ indicated above, correspond
to flux ratios more than an order of magnitude larger than those corresponding to the
typical value of $q\approx 2.3$ found for star forming galaxies
\cite[e.g.,][]{Con:92}. Our estimates are, however, consistent with the elevated values of $q$ identified
within nearby galaxies in regions of elevated star formation \citep{Hug:06,Mur:06}.

\begin{figure} [h]
	\begin{center}
		\includegraphics[angle=-90,scale=0.3,keepaspectratio]{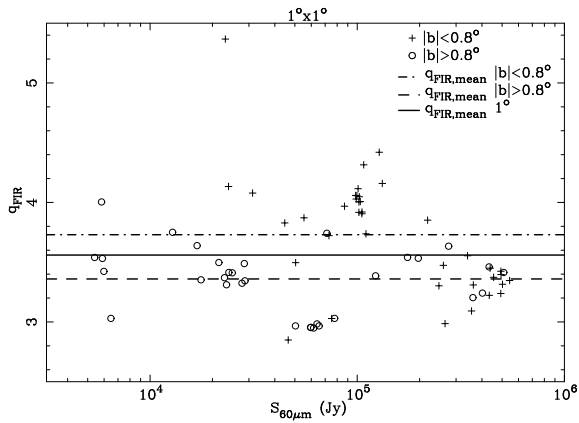} 
		\caption {Values of $q$ as a function of $S_{\rm 60\mu m}$ for the two subsamples
		split by distance from the Galactic plane. $q_{\rm mean}$ is elevated closer to the
		Galactic plane, where the star formation, indicated by $S_{\rm 60\mu m}$, is most active.
		\label{fig3}}
	\end{center}
	\end {figure}

To measure any possible galactic latitude dependence of $q_{\rm mean}$, we use the
$1^{\circ}$ images, although the results are similar with other field sizes. The images were
divided into two bins of galactic latitude, $|b|\leq 0.8^\circ$ and $|b|>0.8^\circ$, containing
39 and 32 target regions respectively. The value of $q_{\rm mean}$ shows a clear galactic latitude
dependence, having higher values on average closer to the galactic plane (Figure~\ref{fig3}). A
Kolmogorov-Smirnov (KS) test comparing the $q_{\rm mean}$ values for the two subsamples gives
$P<1\times10^{-3}$, indicating that they are inconsistent with having been drawn from the same
population. The $S_{\rm 60\mu m}$ flux values for each of these subsamples were also compared
using a KS test. This shows again that, with $P<1\times10^{-3}$, the $60\,\mu$m fluxes of the
subsamples are inconsistent with being drawn from the same population, and that those at lower
latitudes show higher $60\,\mu$m fluxes consistent with having higher star formation rates.

\section{Discussion}
The first main result of this investigation is that a local radio-FIR correlation seems to arise at spatial
scales of at least $\approx 20-60\,$pc, and likely on scales as small as a few pc. The second
main result is that the FIR-radio ratio ($q_{\rm mean}$) has a galactic latitude
dependence. We interpret this second result to indicate $q$ is dependent on the level of
star formation activity.

The scale on which the radio-FIR correlation appears around star formation regions is consistent with
the LMC study \citep{Hug:06} where the correlation arises at $\approx20\,$pc and becomes tight
above $\approx50\,$pc. We further find a correlation on even smaller scales, down to $\approx 4\,$pc.
The diffusion length of relativistic electrons contributing to the galactic
FIR-radio ratio \cite[$\approx1-2\,$kpc,][]{Bic:90} is much larger than these spatial scales. This suggests that the correlation on such a local scale is between thermal radio and FIR, an interpretation consistent with other studies showing strong correlation between thermal radio and warm dust emission
\citep{Hoe:98,Hug:06}. The thermal fraction at 843\,MHz calculated from Equation~5 of \cite{Con:92} is
about $0.08$, so assuming that the FIR emission is constrained to the locale of the star forming region,
the flux ratios observed would be expected to be about an order of magnitude larger than those
corresponding to $q\approx 2.3$ found as the average for star forming galaxies in general. This is
consistent with the
values that we have measured, supporting the result that the high values of $q_{\rm mean}$ from this
study, as well as earlier estimates \citep{Bou:88}, are a consequence of not including
the synchrotron radio emission that is distributed on larger scales.

The result of the latitude dependence analysis is consistent with a dependence of $q_{\rm mean}$ on
the star formation rate.
This is based on the assumption that star formation rates are higher closer to the galactic plane,
supported by the increased $S_{\rm 60\mu m}$ measured there. Our
results are consistent with those of \cite{Hug:06}, who observed elevated $q$ values in the bar of the
LMC where there is elevated star formation activity. Strong additional support for our intepretation of the
galactic latitude analysis comes from a study of local spiral galaxies. \cite{Mur:06} observed that active
star forming regions of galaxies with high SFR display elevated $q$ values.

From Table~\ref{tab1}, the change in $q_{\rm mean}$ between the different spatial scales is unlikely to
be statistically significant given the rms scatter in the measurements. Even out to scales of $2^{\circ}$
(spatial scales of up to $\approx 200\,$pc), where these were able to be measured, the
value of $q_{\rm mean}$ remains essentially unchanged, although there is perhaps
a very marginal elevation on scales of $30'$ to $1^{\circ}$. This, while not statistically significant, may imply a suggestive trend. We suggest that a rise in $q_{\rm mean}$ at $\approx20-50\,$pc and the
decrease on larger scales around star forming regions could be due to the various components of FIR
and radio emission occuring on different spatial scales. The
change in $q_{\rm mean}$ as spatial scale increases is then a consequence of the changing
proportions of these different components. We can describe such a behaviour with a simple
model for the distribution of these components consistent with the known spatial scales of each
emission process.

	\begin{figure} [h]
	\begin{center}
		\includegraphics[angle=-90,scale=0.3,keepaspectratio]{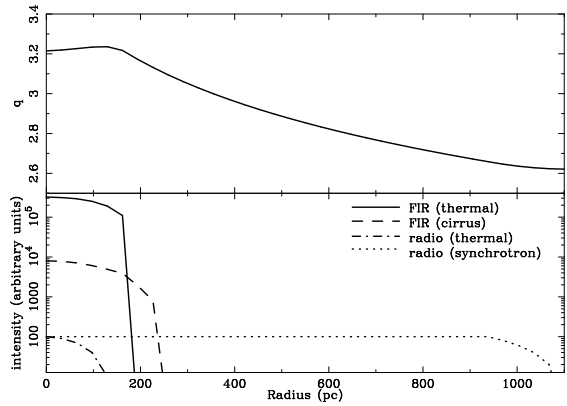}
		\caption {A simple model of the radial scale on which the components of FIR and radio
		emission extend, centred on a star forming region.
		\label{fig4}}
	\end{center}
	\end {figure}

On the smallest scale, thermal FIR and radio are dominant. The thermal radio emission is due to
free-free emission in ionised gas while the thermal FIR emission is from hot dust.
On a slightly larger scale, cirrus FIR emission continues to contribute, eventually ceasing and leaving
only the synchrotron radiation component on the largest scale.

The simple model shown in Figure~\ref{fig4} demonstrates that any rise in $q_{\rm mean}$ between
the $30'$ and the $1^{\circ}$ images could be due to the cirrus FIR emission continuing to greater
spatial scales than the thermal radio emission. The value of q, of course, is calculated using fluxes
corresponding to the integral of these intensities within a given scale. On the kiloparsec scale, and for
a single star forming region, $q$ would continue to decrease with the increasing contribution of the
synchrotron radiation, since all the FIR emission has ceased. Within galaxies, the overlapping of
adjacent star forming regions on scales comparable to the synchrotron diffusion length
then gives rise to the observed lower average value of $q$ dominated by the synchrotron component.

This model, focussing on the dependence of $q$ on spatial scale, does not directly address
whether or not $q$ should be higher around a region of higher star formation rate, for the same
spatial scale sampled (Figure~\ref{fig3}). This may be the case, for example, if the relative proportion of
thermal FIR and radio emission is itself dependent on the star formation rate.

\section{Conclusion}
We have examined the scale on which the radio-FIR correlation arises around star forming regions in
the Milky Way. The correlation is tight by scales of $\approx~20-60\,$pc, and exists on scales even
as small as $\approx 4\,$pc. The galactic latitude
dependence of $q_{\rm mean}$ is consistent with that found in the LMC and local spiral galaxies
\citep{Hug:06,Mur:06}, and strongly suggests that $q$ is elevated in more active star forming
regions. We interpret this as a correlation between the thermal radio emission and the FIR.

We have suggested a simple model that accounts for these observed trends in $q_{\rm mean}$. To test
this model, we plan to compare individual $q$ values, on a range of spatial scales, with the directly
measured star formation
rates in each CHaMP target region. Star formation rates can be measured in a number of different
ways, including the radio and FIR luminosities measured specifically to encompass each star
formation region. For such radio measurements, which are likely to be primarily thermal emission,
existing star formation rate calibrations for total radio luminosity can be revised by
referring to the thermal to total radio flux ratios of \cite{Con:92}. H$\alpha$ luminosities, as well,
may be able to be obtained for many of the regions, and used in estimating star formation rates.
Additional data for the CHaMP targets is also planned to be obtained through programs on
Gemini-S, SOFIA, and Herschel.

Multifrequency radio continuum measurements for each star forming region would allow determination
of the thermal-to-synch\-ro\-tron ratio as a function of spatial scale. This would provide a more accurate quantitative constraint on the scales within the simple model proposed here. It would also allow a refinement of
the radio-derived star formation rates.\\

We warmly thank the referee for extensive and thorough comments that have led to significant
improvements in this paper. AMH acknowledges support provided by the Australian Research
Council through a QEII Fellowship (DP0557850). PJB acknowledges support through NSF
grant AST-0645412 at the University of Florida.

\end{document}